\documentstyle[twocolumn,pre,aps]{revtex}
\begin{document}
\renewcommand{\vec}[1]{{\bf #1}}
\newcommand{\vecg}[1]{\mbox{\boldmath $#1$ }}
\newcommand{\Hbar}{{\bf \overline{H}}}
\newcommand{\Hright}{\Hbar}
\newcommand{\Hleft}{\Hbar}

\draft

\title{
Calculating response functions in time domain with non-orthonormal basis sets
}
\author{ Toshiaki Iitaka 
and Toshikazu Ebisuzaki }
\address{
Division of Computational Science \\
RIKEN (The Institute of Physical and Chemical Research) \\
2-1 Hirosawa, Wako, Saitama 351-0198, JAPAN 
}

\date{\today}

\maketitle

\begin{abstract}
We extend the recently proposed order-$N$ algorithms for calculating linear- and nonlinear-response functions in time domain to the systems described by nonorthonormal basis sets.
\end{abstract}

\pacs{02.70.-c, 02.70.Lq, 31.15.Ar}

\section{Introduction}

As first-principles calculations become more and more important in various research fields such as physics, chemistry, materials science, and recently geology and biology, the demand for calculation of larger and larger systems is growing rapidly.
One of the answers to this demand is the so-called order-$N$ methods, which compute the electronic band structure, the total energy, and other quantities with computational time and storage proportional to N, the number of the atoms in the system.  
For very large systems, these methods are much faster than the conventional diagonalization methods, which require computational efforts proportional to $N^3$.  

The order-$N$ methods may be classified into two steps.  The first step is minimizing the total energy to obtain the ground state of the self-consistent one-particle Hamiltonian. The second step is extracting dynamic properties such as linear and nonlinear-response functions from this Hamiltonian.
While the first step has been extensively studied 
\cite{Galli92,Li93,Hernandez95,Mauri93,Ordejon95,Kim95,Goedecker94,Goedecker95} and also comprehensive reviews are available \cite{Ordejon98,Goedecker99}, the second step has been studied by only few papers \cite{Voter96,Silver96,Roeder97,Wang94,Hobbs96},
including the {\it particle source method}  \cite{Iitaka95,Tanaka98} and the {\it projection method} \cite{Iitaka97,Nomura97,Iitaka99,Kurokawa99}, which use the numerical solution of the time-dependent Schr\"{o}dinger equation \cite{Iitaka94}, and  {\it projected random vectors} \cite{Sankey94}. 

The purpose of this Rapid Communication is to extend the formalism of the projection method to nonorthonormal basis sets\cite{Ballentine86,Gibson93,Stephan98,Goedecker95a,Tsuchida96}, on which many order-$N$ total energy minimization methods are built, so that the full {\it ab initio} calculation from the total energy minimization to the response function is possible.

\section{nonorthonormal basis set}

In this section, let us review the description of a system with a Hilbert space spanned by finite numbers of linearly independent nonorthonormal bases $\{ |\varphi_\alpha \rangle \}$.
We distinguish a vector in the Hilbert space from its components
by using the braket notation for a vector in the Hilbert space 
and the tensor notation \cite{Ballentine86} and the matrix notation \cite{Stephan98} for its components. 

The overlap matrix is defined as a Hermitian matrix with subscripts,
\begin{equation}
S_{\alpha\beta} \equiv \langle \varphi_\alpha | \varphi_\beta \rangle= S_{\beta\alpha}^{*}
.
\end{equation}
Then the inverse matrix is defined as a matrix with superscripts 
that satisfies
\begin{equation}
\sum_\beta \left( S^{-1} \right)^{\alpha\beta}S_{\beta\gamma}
=\delta^{\alpha}_{\ \gamma},
\end{equation}
where $\delta^{\alpha}_{\ \gamma}$ is Kroneker's delta.
Then the {\it dual} basis set $\langle \varphi^{\alpha} |$ is defined by
\begin{equation}
\label{eq:dual.def}
|\varphi^{\alpha}\rangle 
= \sum_\beta |\varphi_{\beta}\rangle \left( S^{-1} \right)^{\beta\alpha},
\end{equation}
which is used only in formal description, but not in real numerical calculations.
These two basis sets are {\it biorthogonal}
and {\it bicomplete},
\begin{eqnarray}
\langle \varphi^{\alpha}| \varphi_{\beta} \rangle
&=& \sum_\gamma (S^{-1})^{\alpha\gamma}S_{\gamma\beta}
= \delta^{\alpha}_{\beta}, \\
\sum_\alpha | \varphi_{\alpha} \rangle \langle \varphi^{\alpha}| 
&=& I,
\end{eqnarray}
where $I$ is the identity operator.

An arbitrary state $|\phi\rangle$ can be expressed in original or dual basis set,
\begin{eqnarray}
\label{eq:phi.expand}
|\phi\rangle &=& \sum_\alpha \phi^{\alpha} | \varphi_\alpha \rangle
             = \sum_{\alpha,\beta} \phi^{\alpha} 
|\varphi^\beta\rangle  S_{\beta\alpha} 
             = \sum_\beta \phi_{\beta} | \varphi^\beta \rangle ,
\end{eqnarray}
where $\phi^\alpha$ and $\phi_\alpha$ are the components in each basis set, which are related to each other by
\begin{equation}
\label{eq:dual}
\phi_\beta = \sum_\beta S_{\beta\alpha} \phi^{\alpha}
.
\end{equation}
The components of $|\phi\rangle$ are represented by a column vector 
$ 
\vecg{\phi}=\left[ \phi^1, \phi^2, \cdots, \phi^N\right]^t
$, where $t$ indicates the transpose of a vector or matrix,
and its dual $\langle \tilde{\phi}|$ is represented by a row vector 
$
\tilde{\vecg{\phi}}=\left[ \phi_{1}^*, \phi_{2}^*, \cdots, \phi_{N}^*\right]$ .

The lower-indexed components of an operator, the Hamiltonian $\hat{H}$ for example, are defined in the original basis set by
\begin{equation} 
H_{\alpha\beta}=\langle \varphi_\alpha | \hat{H} | \varphi_\beta \rangle.
\end{equation}
Then the mixed-indexed components are defined by
\begin{equation}
\label{eq:H.mixed}
H^\alpha_{\ \beta} = \langle \varphi^\alpha| \hat{H} | \varphi_\beta \rangle
= \sum_\gamma (S^{-1})^{\alpha\gamma}H_{\gamma\beta}
.
\end{equation}
The manipulation of state vectors and operators is most conveniently
expressed in the mixed representation. For example,  
$|\psi\rangle =\hat{H}|\phi\rangle$
becomes $\psi^\alpha = \sum_\beta H^{\alpha}_{\ \beta} \phi^\beta$.
Therefore, we can introduce the matrix notation, 
$
\vecg{\psi}=\Hright \vecg{\phi}
$ 
where the bar over the matrix symbol indicates the raise of the first index $\Hbar=\{H^\alpha_{\ \beta} \}$.
Then Eq.~(\ref{eq:H.mixed}) is rewritten as
\begin{equation}
\Hright={\bf S}^{-1} {\bf H},
\end{equation}
where ${\bf H}$ is the matrix $\{ H_{\alpha\beta} \}$.
Now $\Hright$ is not Hermitian matrix anymore, since
\begin{eqnarray}
\Hright^\dagger 
&=& \left( {\bf S}^{-1} {\bf H} \right)^\dagger
=  {\bf H}^\dagger \left( {\bf S}^{-1} \right)^\dagger \\
&=&  {\bf H} {\bf S}^{-1} = {\bf S} \Hbar {\bf S}^{-1} \ne \Hright 
.
\end{eqnarray}
Note that the full calculation of ${\bf S}^{-1}$, which costs $O(N^3)$ CPU time, is not necessary to obtain a good approximant of $\Hbar$ from a sparse ${\bf H}$ \cite{Gibson93,Stephan98}. 
One of the advantages of $\Hright$ over ${\bf H}$ is that power of $\hat{H}$ is easily calculated without explicitly multiplying ${\bf S}^{-1}$ \cite{Stephan98},
\begin{eqnarray}
 \hat{H}^n |\phi\rangle &=& \sum_\beta \hat{H}^n |\varphi_\beta\rangle \phi^\beta
= \sum_\alpha \sum_\beta  |\varphi_\alpha\rangle (\Hright^n)^\alpha_{\ \beta} \phi^\beta \\
\label{eq:Hbar.multiply}
&=& \sum_\alpha |\varphi_\alpha\rangle \left( \Hright^n \vecg{\phi} \right)^\alpha
.
\end{eqnarray}

The matrix form of the eigenvalue problem
\begin{equation}
\sum_\beta H^\alpha_{\ \beta} \phi^\beta = E \phi^\alpha
\end{equation}
becomes
\begin{equation}
\label{eq:eigen.ket.matrix}
\Hright \vecg{\phi}(E_\beta)= E_\beta \vecg{\phi}(E_\beta)
\end{equation}
and the dual of Eq.~(\ref{eq:eigen.ket.matrix}) becomes
\begin{equation}
\label{eq:eigen.bra.matrix}
\tilde{\vecg{\phi}}(E_\beta) \Hleft = E_\beta \tilde{\vecg{\phi}}(E_\beta) 
.
\end{equation}
The eigenvectors, Eqs.~(\ref{eq:eigen.ket.matrix}) and (\ref{eq:eigen.bra.matrix}), define the eigenstates
\begin{eqnarray}
|E_\beta \rangle &=& \sum_\alpha |\varphi_\alpha \rangle \phi^\alpha(E_\beta), \\
\langle \tilde{E_\beta} | &=& \sum_\alpha \tilde{\phi}_{\alpha} (E_\beta)\langle \varphi^\alpha |,
\end{eqnarray}
which satisfy the biorthonormality 
and the bicompleteness
\begin{eqnarray}
\langle \tilde{E_\alpha} | E_\beta \rangle &=& \delta_{\alpha\beta}, \\
\label{eq:eigen.complete}
\sum_\alpha |E_\alpha \rangle \langle \tilde{E_\alpha} | &=& I
.
\end{eqnarray}

\section{Random vectors}
Let us define {\it random states}\cite{Skilling89,Drabold93} by 
\begin{eqnarray}
\label{eq:randomvec.1}
|\Phi \rangle &\equiv& 
\sum_{\beta} |\varphi_\beta \rangle \Phi^\beta , \\
\label{eq:randomvec.2}
\langle \tilde{\Phi} |  &\equiv& 
\sum_{\alpha}  \tilde{\Phi}_\alpha \langle \varphi^\alpha | ,
\end{eqnarray}
where $\{ |\varphi_\beta \rangle \}$ and $\{ \langle \varphi^\alpha | \}$
 are the basis set used in the computation and its dual basis set, 
 respectively.

Their components
\begin{equation}
\label{eq:rondomvec.components}
\Phi^\alpha=\tilde{\Phi}_\alpha^*=\xi_\alpha 
\end{equation}
are the pseudorandom numbers that satisfy  the statistical relation 
\begin{equation}
\left\langle \left\langle \  \xi_{\alpha}^*  \xi_{\beta} \  \right\rangle \right\rangle = \delta_{\alpha\beta}
\end{equation}
where $\left\langle \left\langle \   \cdot \  \right\rangle \right\rangle \ $ indicates the statistical average. 
Note that the transformation of the random vector to its dual does not contain the overlap matrix ${\bf S}$ in Eq.~(\ref{eq:rondomvec.components}), unlike the general rule for usual vectors in Eq.~(\ref{eq:dual}).

These random vectors may be also expressed by the eigenstates of $\Hbar$ by substituting Eq.~(\ref{eq:eigen.complete}) into Eqs.~(\ref{eq:randomvec.1}) and (\ref{eq:randomvec.2}),
\begin{eqnarray}
|\Phi \rangle &=& 
\sum_{\beta\gamma} | E_\beta \rangle
\langle \tilde{E}_\beta |\varphi_\gamma \rangle \xi_\gamma
=\sum_{\beta} | E_\beta \rangle \zeta_\beta, \\
\langle \tilde{\Phi} |  &=&  
\sum_{\alpha\delta} \xi_\delta^{*} 
\langle \varphi^\delta |E_\alpha \rangle \langle \tilde{E}_\alpha |
=\sum_{\alpha}  \zeta_\alpha^* \langle \tilde{E}_\alpha | ,
\end{eqnarray}
where
\begin{eqnarray}
\zeta_\beta &=& \sum_\gamma \langle \tilde{E}_\beta |\varphi_\gamma \rangle \xi_\gamma , \\
\zeta_\alpha^{*} &=& \sum_{\delta} \xi_\delta^{*} \langle \varphi^\delta |E_\alpha \rangle
.
\end{eqnarray}
Although we do not know the actual value of $\zeta_\alpha^*$, $\zeta_\beta$, $\langle \tilde{E}_\beta |$, or $ | E_\alpha \rangle $, we can derive the statistical relation of the random variables $\zeta_\beta$ as follows:
\begin{eqnarray}
\label{eq:stat.zeta}
\left\langle \left\langle \  \zeta_{\alpha}^*  \zeta_{\beta} \  \right\rangle \right\rangle 
&=& \sum_{\gamma}  \sum_{\delta}
\langle \varphi^\delta |E_\alpha \rangle \langle \tilde{E}_\beta | \varphi_\gamma  \rangle
\left\langle \left\langle \  \xi_{\delta}^*  \xi_{\gamma} \  \right\rangle \right\rangle \nonumber \\
&=& \sum_{\gamma} \langle \tilde{E}_\beta | \varphi_\gamma \rangle \langle \varphi^\gamma |E_\alpha \rangle 
= \langle \tilde{E}_\beta | E_\alpha \rangle 
= \delta_{\alpha\beta}
.
\end{eqnarray}
This relation is very important, as we will see later.

One of the useful features of random states is that 
the expectation value of an operator $\hat{X}$ in terms of
the random states gives trace of the operator,
\[
\left\langle \left\langle \ \langle \tilde{\Phi} | \hat{X} | \Phi \rangle \ \right\rangle \right\rangle 
= \sum_{\alpha,\beta} \left\langle \left\langle  \xi_\alpha^* \xi_\beta  \right\rangle \right\rangle 
\langle \varphi^\alpha| \hat{X} | \varphi_\beta \rangle
= \sum_{\alpha} 
X^\alpha_{\ \alpha}
\]
which is identical to the trace calculated with an orthonormal basis set $|n\rangle $ because
\begin{eqnarray}
\left\langle \hat{X} \right\rangle &=& {\rm tr} [\hat{X}] 
= \sum_{n,\alpha,\beta} \langle n | \varphi_\alpha \rangle 
X^\alpha_{\ \beta}
\langle \varphi^\beta| n \rangle \\
&=& \sum_{\alpha,\beta} \langle \varphi^\beta | \varphi_\alpha \rangle 
X^\alpha_{\ \beta} = \sum_{\alpha} 
X^\alpha_{\ \alpha}
.
\end{eqnarray}

\section{Projected Random Vectors}
Then the projected random vectors are defined by
\begin{eqnarray}
\label{eq:step.def1}
\vecg{\Phi_{E_f}} &=& \theta(E_f-\Hbar) \vecg{\Phi} 
=\sum_m c_m \vecg{\psi}_m \\
\label{eq:step.def2}
\tilde{\vecg{\Phi}}_{\!\!\!E_f} &=& \tilde{\vecg{\Phi}} \theta(E_f-\Hbar) 
=\sum_m c_m \tilde{\vecg{\psi}}_m
\end{eqnarray}
where 
$c_m$ are the coefficients for the Chebyshev polynomial expansion of the step function \cite{Silver96,Recipes}
\begin{equation}
 \theta(x)= \left\{
 \begin{array}{c}
 $0$ \ \ \ (\mbox{$x < 0$}) \\
 $1$ \ \ \ (\mbox{$x > 0$}) .
 \end{array}
 \right.
\end{equation}
The random vectors multiplied by the Chebyshev polynomial $T_m(\Hbar)$ 
\begin{eqnarray}
\vecg{\psi}_m &=& T_m(\Hbar) \vecg{\Phi} , \\
\tilde{\vecg{\psi}}_m &=& \tilde{\vecg{\Phi}} T_m(\Hbar) ,
\end{eqnarray}
 are calculated by using the recursion formulas
\begin{eqnarray}
\vecg{\psi}_{m+1} &=& 2\Hbar \vecg{\psi}_m-\vecg{\psi}_{m-1} , \\
\tilde{\vecg{\psi}}_{m+1} &=&  2 \tilde{\vecg{\psi}}_m \Hbar 
- \tilde{\vecg{\psi}}_{m-1} .
\end{eqnarray}

The coefficient vectors, Eqs.(\ref{eq:step.def1}) and ({eq:step.def2}), define the projected random states
\begin{eqnarray}
|\Phi_{E_f}\rangle &\equiv& 
\sum_\alpha  |\varphi_\alpha\rangle \left(\Phi_{E_f}\right)^\alpha 
=\sum_{E_\beta \le E_f}  |E_\beta\rangle \zeta_\beta  , \\
\langle \tilde{\Phi}_{E_f}| &\equiv& 
\sum_\alpha \left(\tilde{\Phi}_{E_f}\right)_\alpha 
\langle \varphi^\alpha | 
= \sum_{E_\beta \le E_f} \zeta^*_\beta \langle \tilde{E}_\beta | 
.
\end{eqnarray}
One of the useful features of projected random states is that 
the expectation value of an operator $\hat{X}$ with them gives the trace of the operator over the Fermi occupied states,
\begin{eqnarray}
\left\langle \left\langle \ \langle \tilde{\Phi}_{E_f} | \hat{X} | \Phi_{E_f} \rangle \ \right\rangle \right\rangle 
&=& \sum_{E_\alpha,E_\beta \ \le E_f} \left\langle \left\langle  \zeta_\alpha^* \zeta_\beta \right\rangle \right\rangle
\langle \tilde{E}_\alpha| \hat{X} | \tilde{E}_\beta \rangle \\
&=& \sum_{E_\alpha \le E_f} 
X^\alpha_{\ \alpha} ,
\end{eqnarray}
where the statistical relation Eq.~(\ref{eq:stat.zeta}) is used.

\section{Time Evolution}
The time-dependent Schr\"{o}dinger equations corresponding to
the eigenvalue equations (\ref{eq:eigen.ket.matrix}) and (\ref{eq:eigen.bra.matrix}) become
\begin{eqnarray}
+i \frac{d}{dt} \vecg{\phi}(t) &=&
\Hright \vecg{\phi}(t) , \\
-i \frac{d}{dt} \tilde{\vecg{\phi}}(t) &=&
\tilde{\vecg{\phi}}(t) \Hbar 
.
\end{eqnarray}
The formal solutions of the time-dependent equations become
\begin{eqnarray}
\vecg{\phi}(t) &=& e^{-i\Hbar t} \vecg{\phi}(t=0)  , \\
\tilde{\vecg{\phi}}(t) &=& \tilde{\vecg{\phi}}(t=0) e^{+i\Hbar t}  
.
\end{eqnarray}
For numerically calculating the time evolution of the coefficients,
we use the leap frog method \cite{Iitaka94},
\begin{eqnarray}
\vecg{\phi}(t+\Delta t) &=& 
-2i\Delta t \Hbar\vecg{\phi}(t) 
+ \vecg{\phi}(t-\Delta t) , \\
\tilde{\vecg{\phi}}(t+\Delta t) &=& 
+2i\Delta t \tilde{\vecg{\phi}}(t) \Hbar
+ \tilde{\vecg{\phi}}(t-\Delta t) ,  
\end{eqnarray}
where $\Delta t$ is the time step.

\section{Linear Response Function}

When an impulse of perturbation $\hat{A}\delta(t)$ is applied to the system described by the Hamiltonian $\hat{H}$, the time evolution of the wave function is described by the time-dependent Schr\"{o}dinger equation in the matrix form
\begin{eqnarray}
i \frac{d}{dt} \vecg{\Phi}(t) &=& \left\{ \Hbar+\overline{\bf A}\delta(t) \right\} \vecg{\Phi}(t) , \\
-i \frac{d}{dt} \tilde{\vecg{\Phi}}(t) &=&  \tilde{\vecg{\Phi}}(t)  \left\{ \Hbar+\overline{\bf A}\delta(t) \right\} ,
\end{eqnarray}
where $\overline{\bf A}={\bf S}^{-1}{\bf A}$ is the matrix of $\hat{A}$ in the mixed representation.
Note that the impulse $\overline{\bf A}\delta(t)$ contains all frequency components $\overline{\bf A} e^{-i\omega t}$.
Assuming that the system was in a projected random state $\vecg{\Phi}^{(0)}=\vecg{\Phi}_{\!\!\! E_f}$ before the perturbation, the wave function after the perturbation ($t>0$) becomes
\begin{eqnarray}
\vecg{\Phi}(t) &=& \vecg{\Phi}^{(0)}(t) + \delta\vecg{\Phi}(t)  , \\
\tilde{\vecg{\Phi}}(t) &=& \tilde{\vecg{\Phi}}^{(0)}(t) + \delta\tilde{\vecg{\Phi}}(t) ,
\end{eqnarray}
where
\begin{eqnarray}
\vecg{\Phi}^{(0)}(t) &=& e^{-i\Hbar t}\vecg{\Phi_{E_f}} , \\
\label{eq:perturb1}
\delta\vecg{\Phi}(t) &=& (-i)e^{-i\Hbar t} \theta(\Hbar-E_f) \overline{\bf A}\vecg{\Phi_{E_f}} ,
\end{eqnarray}
and 
\begin{eqnarray}
\tilde{\vecg{\Phi}}^{(0)}(t) &=& \tilde{\vecg{\Phi}}_{\!\!\!E_f} e^{+i\Hbar t} , \\
\label{eq:perturb2}
\delta\tilde{\vecg{\Phi}}(t) &=& (+i)\tilde{\vecg{\Phi}}_{\!\!\!E_f} \overline{\bf A} \theta(\Hbar-E_f) e^{+i\Hbar t}
\end{eqnarray}
are the time evolution of unperturbed and perturbed vectors.
In Eqs.~(\ref{eq:perturb1}) and (\ref{eq:perturb2}), projection operators $\theta(\overline{\bf H}-E_f)$ have been introduced to ensure that the excited states should be higher than the Fermi energy.

The linear response of an observable $\hat{B}$ from all electrons is calculated as
\begin{equation}
\label{eq:correlation}
\!\!\! \!\!\!
\delta B(t)= 2 \ {\rm Re} \left\{
\delta\tilde{\vecg{\Phi}}_{\!\!\!E_f}(t)
\overline{\bf B} 
\vecg{\Phi}_{\!\!\!E_f}^{(0)}(t)
\right\} , 
\end{equation}
where $\overline{\bf B}={\bf S}^{-1}{\bf B}$ is the matrix of $\hat{B}$ in the mixed representation.
Then the Fourier transformation of $\delta B(t)$ gives the linear response of the noninteracting many-electron system to the perturbation $\overline{\bf A} e^{-i\omega t}$ ,
\begin{eqnarray}
\label{eq:chi.time.one}
\!\!\! \!\!\!
\chi_{BA}(\omega+i\eta) 
&=& 
\label{eq:chi.numerical.1}
\left\langle \!\!\! \left\langle
\rule{0pt}{24pt}
\int_0^T \!\!\!dt \ \ e^{+i(\omega + i\eta)t}  \delta B(t)
\right\rangle \!\!\! \right\rangle ,
\end{eqnarray}
where the imaginary part of frequency $\eta$ is introduced to limit the integration time to a finite value $T = -\ln \delta/\eta $, with $\delta$ being the relative numerical accuracy of Eq.~(\ref{eq:chi.numerical.1}). 
Here $\left\langle \left\langle \   \cdot \  \right\rangle \right\rangle \ $ indicates the statistical average. 

\section{Summary}

We presented a generalized version of the projection 
method for linear and nonlinear response functions developed
by Iitaka and others \cite{Iitaka97,Nomura97,Iitaka99,Kurokawa99}. 
The method can now be used with nonorthonormal basis sets
such as local basis sets, for order-$N$ total energy calculations.
As a result, it became possible to calculate the response 
functions of very large systems by applying the projection
method to the optimized Hamiltonian with a local nonorthonormal
basis set.

\section*{Acknowledgments}

One of the authors (T.I.) would like to thank Professor Ordej\'on 
for useful discussions at the RIKEN Symposium.




\end{document}